\documentstyle[aps,prd,epsfig]{revtex}
\begin{document}
\title{
Acoustic Signatures in the Primary Microwave Background Bispectrum
}

\author{
Eiichiro Komatsu\footnotemark[1]
and
David N. Spergel\footnotemark[2]
}
\address{
Department of Astrophysical Sciences, Princeton University,
Princeton, NJ 08544, USA.
}
\maketitle
\begin{abstract}
If the primordial fluctuations are non-Gaussian,
then this non-Gaussianity will be apparent in the 
cosmic microwave background (CMB) sky.
With their sensitive all-sky observation,
MAP and Planck satellites 
should be able to detect weak non-Gaussianity in the CMB sky.
On large angular scale, there is a simple relationship between
the CMB temperature and the primordial curvature perturbation:
$\Delta T/T=-\Phi/3$.
On smaller scales; however, the radiation transfer function 
becomes more complex.
In this paper, we present  the 
angular bispectrum of the primary CMB anisotropy that 
uses the full transfer function.
We find that the bispectrum has a series of acoustic peaks
that change a sign, and a period of acoustic oscillations is 
twice as long as that of
the angular power spectrum.
Using a single non-linear coupling parameter to characterize  
the amplitude of the bispectrum, 
we estimate the expected 
signal-to-noise ratio
for COBE, MAP, and Planck experiments.
In order to detect the primary CMB bispectrum by each experiment, 
we find that the coupling parameter should be larger than
600, 20, and 5 for COBE, MAP, and Planck experiments, respectively.
Even for the ideal noise-free and infinitesimal thin-beam experiment,
the parameter should be larger than 3. We have included effects from
the cosmic variance, detector noise, and foreground sources in the 
signal-to-noise estimation.
Since the simple inflationary scenarios
predict that the parameter is an order of 0.01,
the detection of the primary bispectrum 
by any kind of experiments should be problematic for those scenarios.
We compare the sensitivity of the primary bispectrum to the primary skewness
and conclude that when we can compute the predicted form of the
bispectrum, it becomes a ``matched filter'' for detecting the 
non-Gaussianity in the data, and much more powerful tool than the 
skewness. For example, we need the coupling parameter of  
larger than 800, 80, 70, and 60 for each relevant experiment
in order to detect the primary skewness.
We also show that MAP and Planck can separate the 
primary bispectrum from various secondary bispectra on the basis 
of the shape difference.
The primary CMB bispectrum is a test of the 
inflationary scenario, and also a probe of the non-linear physics 
in the very early universe.
\end{abstract}

\footnotetext[1]{
komatsu@astro.princeton.edu;
also at the Astronomical Institute, Tohoku University, 
Aoba, Sendai 980-8578, Japan.}
\footnotetext[2]{dns@astro.princeton.edu}

\section{Introduction}

Why measure the bispectrum of the cosmic microwave
background (CMB) radiation anisotropy?
Simple inflationary models predict that the CMB anisotropy 
field is nearly random Gaussian, and that
two-point statistics completely specify statistical properties 
of CMB.
However, our universe may not be so simple.
Higher order statistics, such as the three-point correlation
function, or its harmonic transform, the angular bispectrum, 
are potential probes of the physics of generating
the primordial fluctuations.
Since gravitationally induced non-linearities are small at $z\sim 1300$,
CMB is expected to be the best probe of the primordial
non-Gaussianity\cite{VWHK00}.

In the inflationary scenario\cite{Sato81,Guth81,AS82,Linde82}, 
the quantum fluctuations of the scalar (inflaton)
field generate the observed matter and radiation fluctuations 
in the universe\cite{GP82,Hawking82,BST83,Starobinsky82}. 
In the stochastic inflationary scenario of Starobinsky\cite{Starobinsky86},
the quantum fluctuations decohere to generate the classical fluctuations.
There are two potential sources of non-Gaussianity in this 
inflationary model:
(a) the non-linear coupling between the classical inflaton field and 
the observed fluctuation field, and
(b) the non-linear coupling between the quantum noise field and the
classical fluctuation field.
The former has been investigated by Salopek and Bond\cite{SB9091},
while the latter has been explored by Gangui et al.\cite{GLMM94}.
Calzetta and Hu\cite{CH95} and Matacz\cite{Matacz97} present
an alternative treatment of the decoherence process that leads to 
different results for the primordial density perturbation 
from those obtained by Starobinsky\cite{Starobinsky86}.
Matacz's treatment makes similar predictions for the level of 
non-Gaussianity to 
the Starobinsky's treatment\cite{Matacz97}.
These studies conclude that in the slow roll regime, 
the fluctuations are Gaussian.
However, features in the inflaton potential can produce 
significant non-Gaussianity\cite{KBHP91}.

There have been claims for both the 
non-detection\cite{KBBGHSW96} and the detection\cite{FMG98,Magueijo00}
of the non-Gaussianity in the COBE map.
Banday, Zaroubi and G\'orski\cite{BZG00} argued the non-cosmological 
origin of the COBE non-Gaussianity. 
MAP and Planck will measure the fluctuation field down to 
angular scales $\simeq 0.\hspace{-2.5pt}^\circ 2$ and
$0.\hspace{-2.5pt}^\circ 1$, and test these claims. 

Previous work on the primary non-Gaussianity
has focused on very large angular scale, 
where the temperature fluctuations trace the primordial fluctuations.
This is valid on the COBE scale. For MAP and Planck; however, 
we need the full effect of the radiation transfer function.
In this paper, we develop a formalism for doing this, and then present 
numerical results. Both the formalism and the numerical results are
main results of this paper.
We also discuss how well we can separate the 
primary bispectrum from various secondary bispectra.

This paper is organized as follows. 
Sec.~\ref{sec:formulation} defines the bispectrum, the 
Gaunt integral, and particularly the new quantity called 
the ``reduced'' bispectrum, which plays a fundamental role 
in estimating the physical property of the bispectrum.
Sec.~\ref{sec:bl+s} formulates the primary bispectrum 
that uses the full radiation transfer function, and presents
the numerical results of the primary
bispectrum and the skewness. 
Sec.~\ref{sec:secondary} estimates the secondary bispectra from
the coupling between the Sunyaev--Zel'dovich and the weak lensing
effects\cite{SG99,GS99,CH00}, 
and from the extragalactic radio and infrared sources.
Sec.~\ref{sec:measure} studies how well we can measure each
bispectrum, and how well we can discriminate among various bispectra.
Sec.~\ref{sec:discussion} is devoted to further discussion and 
our conclusion.

\section{Defining the ``reduced'' bispectrum}
\label{sec:formulation}

The observed CMB temperature fluctuation field 
$\Delta T(\hat{\mathbf n})/T$ is expanded into 
the spherical harmonics:
\begin{equation}
  a_{lm}\equiv \int d^2\hat{\mathbf n}\frac{\Delta T(\hat{\mathbf n})}{T}
  Y_{lm}^*(\hat{\mathbf n}),
\end{equation}
where hats denote unit vectors. 
The CMB angular bispectrum is given by 
\begin{equation}
  \label{eq:blllmmm}
  B_{l_1l_2l_3}^{m_1m_2m_3}\equiv 
  \left<a_{l_1m_1}a_{l_2m_2}a_{l_3m_3}\right>,
\end{equation}
and the angle-averaged bispectrum is defined by
\begin{equation}
  \label{eq:blll}
  B_{l_1l_2l_3}\equiv \sum_{m_1m_2m_3}
  \left(
  \begin{array}{ccc}
  l_1&l_2&l_3\\
  m_1&m_2&m_3
  \end{array}
  \right)
  B_{l_1l_2l_3}^{m_1m_2m_3}, 
\end{equation}
where the matrix is the Wigner-$3j$ symbol.
The bispectrum $B_{l_1l_2l_3}^{m_1m_2m_3}$
must satisfy the triangle conditions and 
selection rules: $m_1+m_2+m_3=0$, $l_1+l_2+l_3={\rm even}$, and 
$\left|l_i-l_j\right|\leq l_k \leq l_i+l_j$ for all permutations
of indices. Thus, $B_{l_1l_2l_3}^{m_1m_2m_3}$ consists of
the Gaunt integral,  
${\cal G}_{l_1l_2l_3}^{m_1m_2m_3}$, defined by
\begin{eqnarray}
  \nonumber
  {\cal G}_{l_1l_2l_3}^{m_1m_2m_3}
  &\equiv&
  \int d^2\hat{\mathbf n}
  Y_{l_1m_1}(\hat{\mathbf n})
  Y_{l_2m_2}(\hat{\mathbf n})
  Y_{l_3m_3}(\hat{\mathbf n})\\
  \label{eq:gaunt}
  &=&\sqrt{
   \frac{\left(2l_1+1\right)\left(2l_2+1\right)\left(2l_3+1\right)}
        {4\pi}
        }
  \left(
  \begin{array}{ccc}
  l_1 & l_2 & l_3 \\ 0 & 0 & 0 
  \end{array}
  \right)
  \left(
  \begin{array}{ccc}
  l_1 & l_2 & l_3 \\ m_1 & m_2 & m_3 
  \end{array}
  \right).
\end{eqnarray}
${\cal G}_{l_1l_2l_3}^{m_1m_2m_3}$ is real, and satisfies 
all the conditions mentioned above.

Given the rotational invariance of the universe, 
$B_{l_1l_2l_3}$ is written as
\begin{equation}
  \label{eq:func}
  B_{l_1l_2l_3}^{m_1m_2m_3}
  ={\cal G}_{l_1l_2l_3}^{m_1m_2m_3}b_{l_1l_2l_3}, 
\end{equation}
where $b_{l_1l_2l_3}$ is an 
arbitrary real symmetric function of $l_1$, $l_2$, and $l_3$.
This form of equation (\ref{eq:func}) is necessary and
sufficient to construct generic $B_{l_1l_2l_3}^{m_1m_2m_3}$
under the rotational invariance.
Thus, we shall frequently use $b_{l_1l_2l_3}$ instead of 
$B_{l_1l_2l_3}^{m_1m_2m_3}$ in this paper, and call this function
the ``reduced'' bispectrum, as $b_{l_1l_2l_3}$ 
contains all physical information in $B_{l_1l_2l_3}^{m_1m_2m_3}$.
Since the reduced bispectrum does not contain the Wigner-$3j$ symbol that
merely ensures the triangle conditions 
and selection rules, it is easier to calculate
and useful to quantify the physical properties of the bispectrum.

The observable quantity, the angle-averaged bispectrum $B_{l_1l_2l_3}$,
is obtained by substituting equation (\ref{eq:func}) into (\ref{eq:blll}),
\begin{equation}
  \label{eq:wigner*}
  B_{l_1l_2l_3}
  =
  \sqrt{\frac{(2l_1+1)(2l_2+1)(2l_3+1)}{4\pi}}
  \left(
  \begin{array}{ccc}
  l_1&l_2&l_3\\
  0&0&0
  \end{array}
  \right)b_{l_1l_2l_3},
\end{equation}
where we have used the identity:
\begin{equation}
  \label{eq:wigner}
  \sum_{m_1m_2m_3}
  \left(
  \begin{array}{ccc}
  l_1&l_2&l_3\\
  m_1&m_2&m_3
  \end{array}
  \right)
  {\cal G}_{l_1l_2l_3}^{m_1m_2m_3}
  =
  \sqrt{\frac{(2l_1+1)(2l_2+1)(2l_3+1)}{4\pi}}
  \left(
  \begin{array}{ccc}
  l_1&l_2&l_3\\
  0&0&0
  \end{array}
  \right).
\end{equation}

Alternatively, one can define the bispectrum in the flat-sky
approximation, 
\begin{equation}
 \label{eq:smallangle}
  \left<a({\mathbf l}_1)a({\mathbf l}_1)a({\mathbf l}_3)\right>
  =(2\pi)^2\delta^{(2)}\left({\mathbf l}_1+{\mathbf l}_2+{\mathbf l}_3\right)
  B({\mathbf l}_1,{\mathbf l}_2,{\mathbf l}_3),
\end{equation}
where ${\mathbf l}$ is the two dimensional wave-vector on the sky.
This definition of $B({\mathbf l}_1,{\mathbf l}_2,{\mathbf l}_3)$ 
corresponds to equation (\ref{eq:func}),
given the correspondence of 
${\cal G}_{l_1l_2l_3}^{m_1m_2m_3}\rightarrow 
\delta^{(2)}\left({\mathbf l}_1+{\mathbf l}_2+{\mathbf l}_3\right)$ 
in the flat-sky limit\cite{Hu00}. Thus,
\begin{equation}
 \label{eq:smallangle*}
  b_{l_1l_2l_3}\approx
  B({\mathbf l}_1,{\mathbf l}_2,{\mathbf l}_3)
  \qquad \mbox{(flat-sky approximation)},
\end{equation}
is satisfied. 
This fact also would motivate us to use
the reduced bispectrum $b_{l_1l_2l_3}$ rather than the angular
averaged bispectrum $B_{l_1l_2l_3}$.
Note that $b_{l_1l_2l_3}$ is similar to $\hat{B}_{l_1l_2l_3}$ defined by 
Magueijo\cite{Magueijo00}.
The relation is $b_{l_1l_2l_3}=\sqrt{4\pi}\hat{B}_{l_1l_2l_3}$.

\section{Primary Bispectrum and Skewness}
\label{sec:bl+s}

\subsection{Model of the primordial non-Gaussianity}

If the primordial fluctuations are adiabatic scalar fluctuations, then
\begin{equation}
  \label{eq:almphi}
  a_{lm}=4\pi(-i)^l
  \int\frac{d^3{\mathbf k}}{(2\pi)^3}\Phi({\mathbf k})g_{Tl}(k)
  Y_{lm}^*(\hat{\mathbf k}),
\end{equation}
where $\Phi({\mathbf k})$ is the primordial curvature perturbation
in the Fourier space, and $g_{Tl}(k)$ is the radiation transfer function.
$a_{lm}$ thus takes over the non-Gaussianity,
if any, from $\Phi({\mathbf k})$.
Although equation (\ref{eq:almphi}) is valid only if the
universe is flat, it is straightforward to extend this
to an arbitrary geometry.
The isocurvature fluctuations can be similarly calculated by
using the entropy perturbation and the proper transfer function.

In this paper, we explore the simplest weak non-linear coupling 
case: 
\begin{equation}
  \label{eq:modelreal}
  \Phi({\mathbf x})
 =\Phi_L({\mathbf x})
 +f_{NL}\left(
              \Phi^2_{L}({\mathbf x})-
	      \left<\Phi^2_{L}({\mathbf x})\right>
        \right),
\end{equation}
in real space, where $\Phi_L({\mathbf x})$ denotes the linear gaussian 
part of the perturbation. $\left<\Phi({\mathbf x})\right>=0$ is
guaranteed.
Henceforth, we shall call $f_{NL}$ the non-linear coupling constant.
This model is based upon the slow-roll inflationary 
scenario. Salopek and Bond\cite{SB9091} and Gangui et al.\cite{GLMM94}
found that $f_{NL}$ is given by a certain combination of 
the slope and the curvature of the inflaton potential.
In the notation of Gangui et al., $\Phi_3=2f_{NL}$.
Gangui et al. found that $\Phi_3\sim 10^{-2}$
in the quadratic and the quartic inflaton potential models.

In the Fourier space, $\Phi({\mathbf k})$ is decomposed into two parts:
\begin{equation}
  \label{eq:model}
  \Phi({\mathbf k})=\Phi_L({\mathbf k})+\Phi_{NL}({\mathbf k}),
\end{equation}
and accordingly,
\begin{equation}
  \label{eq:model*}
  a_{lm}=a_{lm}^L+a_{lm}^{NL},
\end{equation}
where $\Phi_{NL}({\mathbf k})$ is the non-linear part defined by
\begin{equation}
  \label{eq:nonlinear}
  \Phi_{NL}({\mathbf k})\equiv 
  f_{NL}
  \left[
  \int \frac{d^3{\mathbf p}}{(2\pi)^3}
  \Phi_L({\mathbf k}+{\mathbf p})\Phi^*_L({\mathbf p})
  -(2\pi)^3\delta^{(3)}({\mathbf k})\left<\Phi^2_{L}({\mathbf x})\right>
  \right].
\end{equation}
One can confirm that $\left<\Phi({\mathbf k})\right>=0$ is
satisfied.
In this model, a non-vanishing component of the 
$\Phi({\mathbf k})$-field bispectrum is
\begin{equation}
  \label{eq:phispec}
  \left<\Phi_L({\mathbf k}_1)
	\Phi_L({\mathbf k}_2)
	\Phi_{NL}({\mathbf k}_3)\right>
  = 2(2\pi)^3\delta^{(3)}({\mathbf k}_1+{\mathbf k}_2+{\mathbf k}_3)
    f_{NL}P_\Phi(k_1)P_\Phi(k_2),
\end{equation}
where $P_\Phi(k)$ is the linear power spectrum given by 
$\left<\Phi_L({\mathbf k}_1)\Phi_L({\mathbf k}_2)\right>
  =(2\pi)^3P_\Phi(k_1)\delta^{(3)}({\mathbf k}_1+{\mathbf k}_2)$.
We have also used 
$\left<\Phi_L({\mathbf k}+{\mathbf p})\Phi^*_L({\mathbf p})\right>
  =(2\pi)^3P_\Phi(p)\delta^{(3)}({\mathbf k})$,
and 
$\left<\Phi^2_{L}({\mathbf x})\right>
=(2\pi)^{-3}\int d^3{\mathbf k} P_\Phi(k)$.

Substituting equation (\ref{eq:almphi}) into (\ref{eq:blllmmm}),
using equation (\ref{eq:phispec}) for the $\Phi({\mathbf k})$-field
bispectrum, and then integrating over angles 
$\hat{\mathbf k}_1$, $\hat{\mathbf k}_3$, and $\hat{\mathbf k}_3$, we obtain 
the primary CMB angular bispectrum,
\begin{eqnarray}
  \nonumber
  B_{l_1l_2l_3}^{m_1m_2m_3}
  &=& 
  \left<a_{l_1m_1}^La_{l_2m_2}^La_{l_3m_3}^{NL}\right>
  + \left<a_{l_1m_1}^La_{l_2m_2}^{NL}a_{l_3m_3}^{L}\right>
  + \left<a_{l_1m_1}^{NL}a_{l_2m_2}^La_{l_3m_3}^{L}\right>\\
  \label{eq:almspec}
  &=& 2{\cal G}_{l_1l_2l_3}^{m_1m_2m_3}
	\int_0^\infty r^2 dr 
    \left[
          b^L_{l_1}(r)b^L_{l_2}(r)b^{NL}_{l_3}(r)+
	  b^L_{l_1}(r)b^{NL}_{l_2}(r)b^{L}_{l_3}(r)+
	  b^{NL}_{l_1}(r)b^L_{l_2}(r)b^{L}_{l_3}(r)
    \right],
\end{eqnarray}
where 
\begin{eqnarray}
  \label{eq:bLr}
  b^L_{l}(r) &\equiv&
  \frac2{\pi}\int_0^\infty k^2 dk P_\Phi(k)g_{Tl}(k)j_l(kr),\\
  \label{eq:bNLr}
  b^{NL}_{l}(r) &\equiv&
  \frac2{\pi}\int_0^\infty k^2 dk f_{NL}g_{Tl}(k)j_l(kr).
\end{eqnarray}
Note that $b^L_{l}(r)$ is a dimensionless quantity, 
while $b^{NL}_{l}(r)$ has a dimension of $L^{-3}$.

One confirms that the form of equation (\ref{eq:func}) holds.
Thus, the reduced bispectrum, 
$b_{l_1l_2l_3}=
B_{l_1l_2l_3}^{m_1m_2m_3}
\left({\cal G}_{l_1l_2l_3}^{m_1m_2m_3}\right)^{-1}$ 
(Eq.(\ref{eq:func})), for the primordial non-Gaussianity is
\begin{eqnarray}
  b_{l_1l_2l_3}^{primary}
  \label{eq:blprim}
  = 2\int_0^\infty r^2 dr 
    \left[
          b^L_{l_1}(r)b^L_{l_2}(r)b^{NL}_{l_3}(r)+
	  b^L_{l_1}(r)b^{NL}_{l_2}(r)b^{L}_{l_3}(r)+
	  b^{NL}_{l_1}(r)b^L_{l_2}(r)b^{L}_{l_3}(r)
    \right].
\end{eqnarray}
$b_{l_1l_2l_3}^{primary}$ is fully specified by a single constant
parameter $f_{NL}$, as the cosmological parameters will be precisely 
determined by measuring the CMB angular power spectrum $C_l$ 
(e.g., \cite{BET97}).
It should be stressed again that this is the special case in the 
slow-roll limit. 
If the slow-roll condition is not satisfied, 
then $f_{NL}=f_{NL}(k_1,k_2,k_3)$ at equation 
(\ref{eq:phispec})\cite{GLMM94}.
Wang and Kamionkowski\cite{WK00} have developed the formula to 
compute $B_{l_1l_2l_3}$ from the generic form of 
$\Phi({\mathbf k})$-field bispectrum.
Our formula (Eq.(\ref{eq:almspec})) agrees with theirs,
given our form of the $\Phi({\mathbf k})$-field bispectrum
(Eq.(\ref{eq:phispec})).

Even if the inflation produced Gaussian fluctuations,
Pyne and Carroll pointed out
that the general relativistic second-order perturbation theory 
would produce terms of $f_{NL}\sim {\cal O}(1)$\cite{PC96}.
For generic slow-roll models, these terms dominate the primary
non-Gaussianity.

\begin{figure}
\begin{center}
    \leavevmode\epsfxsize=9cm \epsfbox{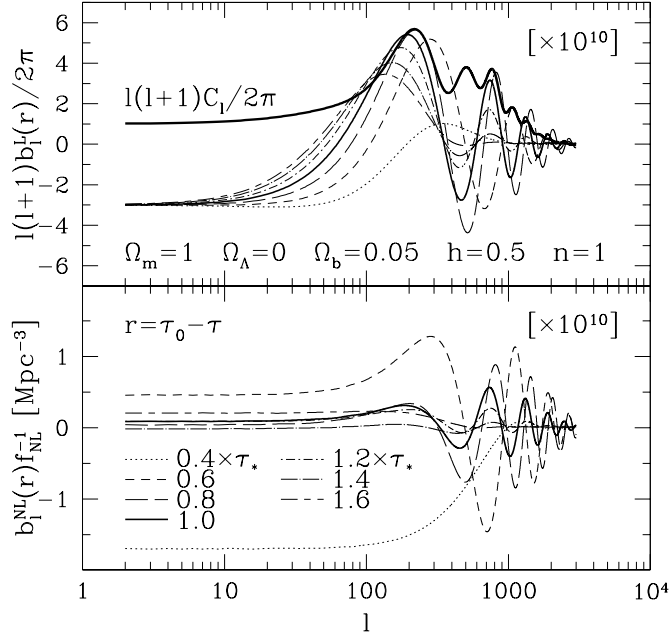}
\end{center}
\caption{
This figure shows $b_l^L(r)$ (Eq.(\ref{eq:bLr})) and 
$b_l^{NL}(r)$ (Eq.(\ref{eq:bNLr})), the two terms in our calculation of 
the primary CMB angular bispectrum, as a function of $r$. 
Various lines in the upper panel show 
$\left[l(l+1)b_l^L(r)/2\pi\right]\times 10^{10}$, where
$r=c\left(\tau_0-\tau\right)$, at $\tau=0.4,0.6,0.8,1.0,1.2,1.4$, 
and $1.6\times \tau_*$ (decoupling time), 
while $\left[b_l^{NL}(r)f^{-1}_{NL}\right]\times 10^{10}$ 
are shown in the lower panel. 
$\tau_0$ is the present-day conformal time.
Note that $c\tau_0=11.8\ {\rm Gpc}$, and $c\tau_*=235\ {\rm Mpc}$ in our 
cosmological model chosen here.
The thickest solid line in the upper panel is the CMB angular power spectrum
$\left[l(l+1)C_l/2\pi\right]\times 10^{10}$.  
$C_l$ is shown for comparison.
}
\label{fig:bl}
\end{figure}

\subsection{Numerical results of the primary bispectrum}


We evaluate the primary CMB bispectrum
(Eqs.(\ref{eq:almspec})--(\ref{eq:blprim})) numerically.
We compute the full radiation transfer function $g_{Tl}(k)$
with the CMBFAST\cite{SZ96} code, and assume
the single power law spectrum, $P_\Phi(k)\propto k^{n-4}$,
for the primordial curvature fluctuations.
The integration over $k$ (Eqs.(\ref{eq:bLr}) and (\ref{eq:bNLr})) 
is done by the algorithm used in CMBFAST.
The cosmological model is the scale-invariant standard 
cold dark matter model with $\Omega_m=1$, $\Omega_\Lambda=0$,
$\Omega_b=0.05$, $h=0.5$, 
and $n=1$, and with the power spectrum $P_\Phi(k)$ 
normalized to COBE\cite{BW97}.
Although this model is almost excluded by current observations,
it is still useful to depict the basic effects of the transfer function 
on the bispectrum.

Figure~\ref{fig:bl} shows $b_l^L(r)$ (Eq.(\ref{eq:bLr})) and 
$b_l^{NL}(r)$ (Eq.(\ref{eq:bNLr}))
for several different values of $r$.
$r=c\left(\tau_0-\tau\right)$, where $\tau$ is the conformal time,
and $\tau_0$ is at the present.
In our model, $c\tau_0=11.8\ {\rm Gpc}$, and 
the decoupling epoch occurs at $c\tau_*=235\ {\rm Mpc}$ 
at which the differential visibility has a maximum.
Our $c\tau_0$ includes the radiation effect on the expansion of
universe, otherwise $c\tau_0=12.0\ {\rm Gpc}$.
$\tau_*$ is the epoch when the most of the primary signal is generated.
$b^L_l(r)$ and $C_l$ look very similar one another in 
the shape and the amplitude at $l\gtrsim 100$, 
although the amplitude in the Sachs--Wolfe regime 
is different by a factor of $-3$. 
This is because $C_l$ is proportional to $P_\Phi(k)g_{Tl}^2(k)$,
while $b_l^L(r)\propto P_\Phi(k)g_{Tl}(k)$, where $g_{Tl}=-1/3$.
$b_l^L(r)$ has a good phase coherence over wide range of $r$,
while the phase of $b_l^{NL}(r)$ in high-$l$ regime 
oscillates rapidly as a function of $r$. 
This strongly damps the integrated result of 
the bispectrum (Eq.(\ref{eq:almspec})) in high-$l$ regime.
The main difference between $C_l$ and $b_l(r)$ is that
$b_l(r)$ changes a sign, while $C_l$ does not.

Looking at figure~\ref{fig:bl}, we find
$l^2b_l^L\sim 2\times 10^{-9}$ and
$b_l^{NL}f^{-1}_{NL}\sim 10^{-10}\ {\rm Mpc^{-3}}$.
The most signal coming from the decoupling, 
the volume element at $\tau_*$ is 
$r_*^2\Delta r_*\sim (10^4)^2\times 10^2\ {\rm Mpc^3}$,
and thus we estimate an order of magnitude of the primary reduced
bispectrum (Eq.(\ref{eq:blprim})) as
\begin{equation}
  \label{eq:orderest}
  b_{lll}^{primary}\sim 
  l^{-4}
  \left[2 r_*^2\Delta r_*\left(l^2b_l^L\right)^2b_l^{NL}\times 3\right]
  \sim l^{-4}\times 2\times 10^{-17}f_{NL}. 
\end{equation}
Since $b_l^{NL}f^{-1}_{NL}\sim r_*^{-2}\delta(r-r_*)$ 
(see Eq.(\ref{eq:deltadelta})),
$r_*^2\Delta r_* b_l^{NL}f^{-1}_{NL}\sim 1$.
This rough estimate agrees with the numerical result below (figure 2).


\begin{figure}
\begin{center}
    \leavevmode\epsfxsize=9cm \epsfbox{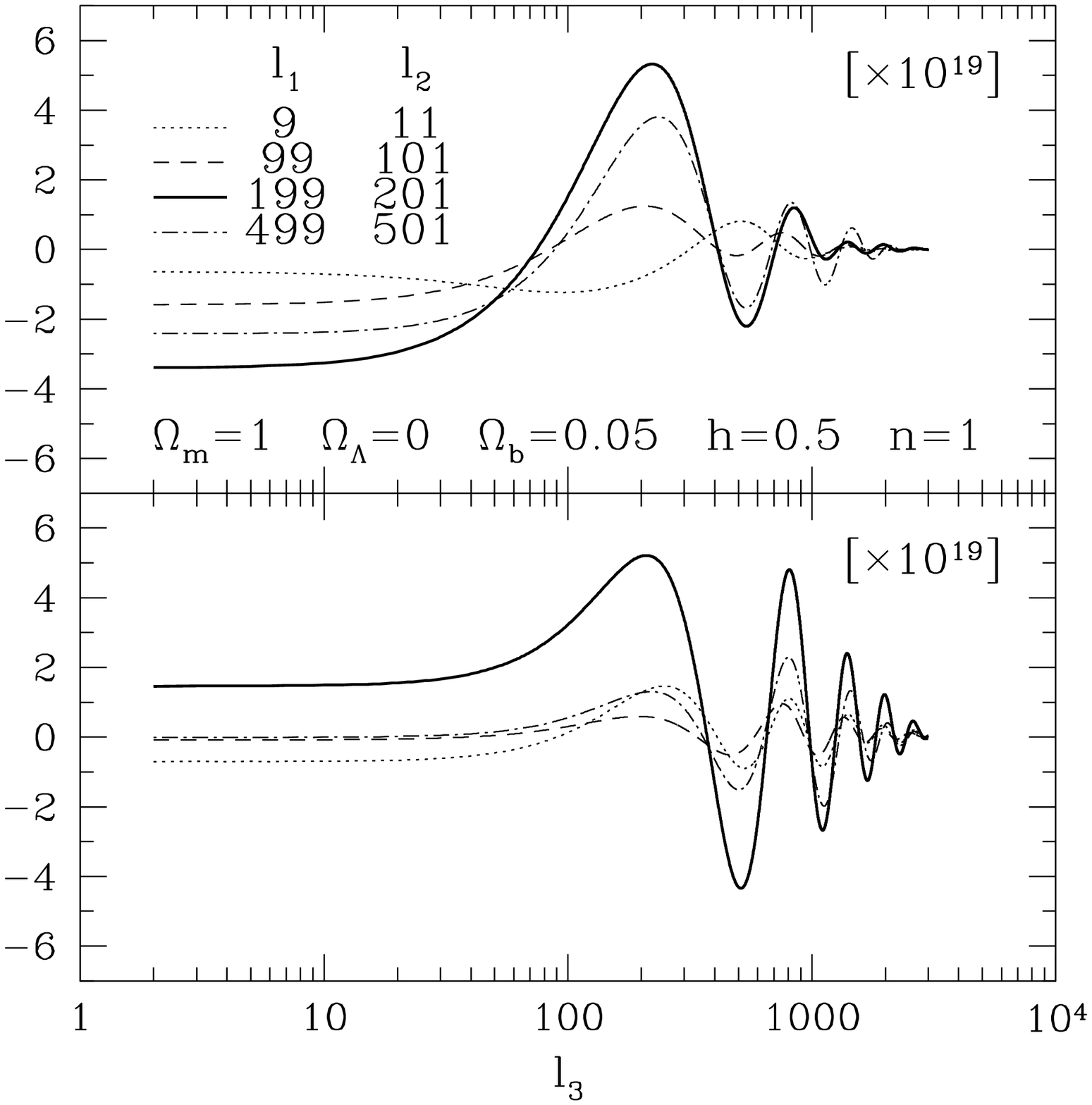}
\end{center}
\caption{
The primary CMB angular bispectrum (Eq.(\ref{eq:almspec})) divided by
the Gaunt integral ${\cal G}_{l_1l_2l_3}^{m_1m_2m_3}$ (Eq.(\ref{eq:gaunt})).
The upper panel shows 
$\left[l_2(l_2+1)l_3(l_3+1)\left<
a_{l_1m_1}^{NL} a_{l_2m_2}^L a_{l_3m_3}^L\right>f_{NL}^{-1}
\left({\cal G}_{l_1l_2l_3}^{m_1m_2m_3}\right)^{-1}/(2\pi)^2\right]
\times 10^{19}$,
while the lower panel shows
$\left[l_1(l_1+1)l_2(l_2+1)\left<
a_{l_1m_1}^{L} a_{l_2m_2}^L a_{l_3m_3}^{NL}\right>f_{NL}^{-1}
\left({\cal G}_{l_1l_2l_3}^{m_1m_2m_3}\right)^{-1}/(2\pi)^2\right]
\times 10^{19}$.
Those are shown as functions of $l_3$ for 
$(l_1,l_2)=(9,11),(99,101),(199,201)$, and $(499,501)$.
}
\label{fig:bispectrum}
\end{figure}

Figure~\ref{fig:bispectrum} 
shows the integrated bispectrum (Eq.(\ref{eq:almspec}))
divided by the Gaunt integral ${\cal G}_{l_1l_2l_3}^{m_1m_2m_3}$,
which is basically $b_{l_1l_2l_3}^{primary}$.
Since the signal comes primarily from the decoupling epoch $\tau_*$ as
mentioned above,
the integration boundary is chosen as 
$c\left(\tau_0-2\tau_*\right)\leq r\leq c\left(\tau_0-0.1\tau_*\right)$.
We use a step-size of $0.1c\tau_*$, as we found that a step size of  
$0.01c\tau_*$ gives very similar results.
While the bispectrum is a 3-d function, we show different 1-d slices of the 
bispectrum in this figure.
$\l_2(l_2+1)l_3(l_3+1)
\left<a_{l_1m_1}^{NL}a_{l_2m_2}^La_{l_3m_3}^L\right>
\left({\cal G}_{l_1l_2l_3}^{m_1m_2m_3}\right)^{-1}/(2\pi)^2$ is
plotted as a function of $l_3$ in the upper panel, while
$l_1(l_1+1)l_2(l_2+1)
\left<a_{l_1m_1}^La_{l_2m_2}^La_{l_3m_3}^{NL}\right>
\left({\cal G}_{l_1l_2l_3}^{m_1m_2m_3}\right)^{-1}/(2\pi)^2$
is plotted in the lower panel.
$l(l+1)/(2\pi)$ is multiplied for each $b_{l}^L(r)$ 
which contains $P_\Phi(k)$ so 
as the Sachs--Wolfe plateau at $l_3\lesssim 10$ is easily seen 
in figure~\ref{fig:bispectrum}.
$l_1$ and $l_2$ are chosen so as $(l_1,l_2)=(9,11),(99,101),(199,201)$,
and $(499,501)$.
We find that the $(l_1,l_2)=(199,201)$ mode,  
the first acoustic peak mode, has the largest signal in this 
family of parameters.
The upper panel has a prominent first acoustic peak, and strongly damped
oscillations in high-$l$ regime.
The lower panel also has a first peak, but damps more slowly.
The typical amplitude of the reduced bispectrum is
$l^4b^{primary}_{lll}f^{-1}_{NL}\sim 10^{-17}$, which agrees with
an order of magnitude estimate (Eq.(\ref{eq:orderest})).

Our formula (Eq.(\ref{eq:blprim})) and numerical results agree with 
Gangui et al.\cite{GLMM94} calculation 
in the Sachs--Wolfe regime, where $g_{Tl}(k)\approx -j_l(kr_*)/3$, and thus
\begin{equation}
  \label{eq:SWapp}
  b_{l_1l_2l_3}^{primary}
  \approx
  -6f_{NL}
  \left(C_{l_1}^{SW}C_{l_2}^{SW}+
        C_{l_1}^{SW}C_{l_3}^{SW}+
        C_{l_2}^{SW}C_{l_3}^{SW}\right)
  \qquad
  \mbox{(Sachs--Wolfe approximation)}.
\end{equation}
Each term is in the same order as equation (\ref{eq:blprim}).
$C_l^{SW}$ is the CMB angular power spectrum in the Sachs--Wolfe
approximation,
\begin{equation}
  \label{eq:clsw}
  C_l^{SW}
  \equiv
  \frac2{9\pi}\int_0^\infty k^2 dk P_\Phi(k)j^2_l(kr_*).
\end{equation}
In deriving equation (\ref{eq:SWapp}) from (\ref{eq:blprim}), 
we approximated $b_l^{NL}(r)$ (Eq.(\ref{eq:bNLr})) to
\begin{equation}
  \label{eq:deltadelta}
  b_l^{NL}(r)
  \approx
  \left(-\frac{f_{NL}}3\right)
  \frac2{\pi}\int_0^\infty k^2 dk j_{l}(kr_*)j_l(kr)
  = -\frac{f_{NL}}3 r_*^{-2}\delta(r-r_*).
\end{equation}
The Sachs--Wolfe approximation (Eq.(\ref{eq:SWapp})) 
is valid only when $l_1$, $l_2$, and $l_3$ are
all less than $\sim 10$, where
Gangui et al.'s formula gives $\sim -6\times 10^{-20}$ in
figure~\ref{fig:bispectrum}.
It should be stressed again that the Sachs--Wolfe approximation 
gives the qualitatively different result from our full calculation 
(Eq.(\ref{eq:blprim})) at
$l_i\gtrsim 10$. The full bispectrum does change a sign, while the 
approximation never changes a sign because of the use of $C_l^{SW}$.
The acoustic oscillation and the sign change 
are actually great advantages, when we try
to separate the primary bispectrum from various secondary bispectra.
We shall study this point later.

\subsection{Primary skewness}


The skewness $S_3$,
\begin{equation}
  S_3\equiv \left<\left(\frac{\Delta T(\hat{\mathbf n})}{T}\right)^3\right>
\end{equation}
is the simplest statistic characterizing the non-Gaussianity. 
$S_3$ is expanded in terms of $B_{l_1l_2l_3}$ (Eq.(\ref{eq:blll}))
or $b_{l_1l_2l_3}$ (Eq.(\ref{eq:func})) as 
\begin{eqnarray}
  \nonumber
  S_3
  &=&
  \frac1{4\pi}\sum_{l_1l_2l_3}
  \sqrt{
  \frac{\left(2l_1+1\right)\left(2l_2+1\right)\left(2l_3+1\right)}
        {4\pi}
        }
  \left(
  \begin{array}{ccc}
  l_1 & l_2 & l_3 \\ 0 & 0 & 0 
  \end{array}
  \right)
   B_{l_1l_2l_3}
   W_{l_1}W_{l_2}W_{l_3}\\
  &=&
  \label{eq:skewness}
  \frac1{2\pi^2}\sum_{2\leq l_1l_2l_3}
  \left(l_1+\frac12\right)\left(l_2+\frac12\right)\left(l_3+\frac12\right)
  \left(
  \begin{array}{ccc}
  l_1 & l_2 & l_3 \\ 0 & 0 & 0 
  \end{array}
  \right)^2
   b_{l_1l_2l_3}
   W_{l_1}W_{l_2}W_{l_3},
\end{eqnarray}
where $W_l$ is the experimental window function. 
We have used equation (\ref{eq:wigner*}) to replace 
$B_{l_1l_2l_3}$ by the reduced bispectrum $b_{l_1l_2l_3}$ 
in the last equality. Since $l=0$ and $1$
modes are not observable, we have excluded them from the summation.
Throughout this paper, we consider the single-beam window function,
$W_l=e^{-l(l+1)/(2\sigma_b^2)}$, where $\sigma_b={\rm FWHM}/\sqrt{8\ln2}$.
Since 
$\left(
\begin{array}{ccc}l_1&l_2&l_3\\0&0&0
\end{array}\right)^2 b_{l_1l_2l_3}$ is symmetric under permutation of
indices, it is useful to change the way of summation as
\begin{equation} 
  \label{eq:sumchange}
  \sum_{2\leq l_1l_2l_3}
  \longrightarrow
  6 \sum_{2\leq l_1\leq l_2\leq l_3}.
\end{equation}
Since this reduces the number of summations by a factor of $\simeq 6$,
we shall use this convention henceforth.

\begin{figure}
\begin{center}
    \leavevmode\epsfxsize=9cm \epsfbox{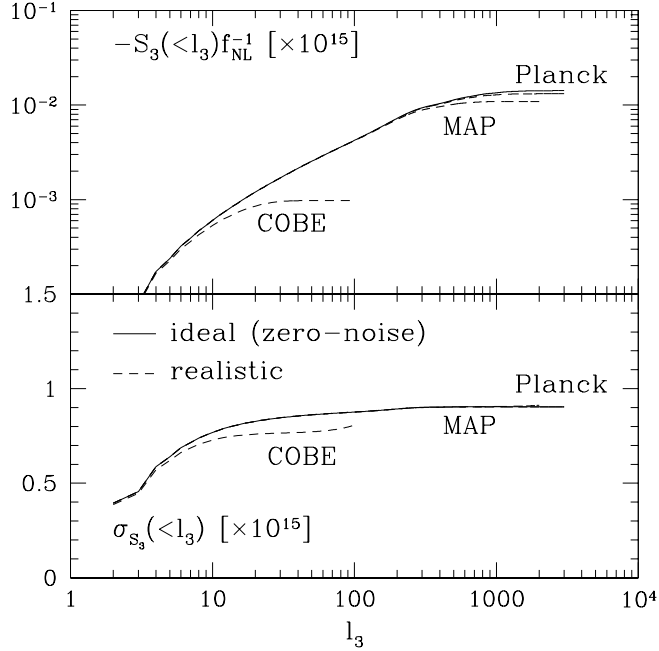}
\end{center}
\caption{
The upper panel shows the primary CMB skewness (Eq.(\ref{eq:skewness})) 
summed up to a certain $l_3$, 
$-S_3(<l_3)f_{NL}^{-1}\times 10^{15}$.
The lower panel shows the noise (Eq.(\ref{eq:skewvar})) 
summed up to $l_3$, $\sigma_{S_3}(<l_3)\times 10^{15}$.
Solid line represents the zero-noise ideal experiment, 
while dotted lines show
COBE, MAP, and Planck experiments.
}
\label{fig:skewness}
\end{figure}

The upper panel of figure~\ref{fig:skewness} plots $S_3(<l_3)$, 
which is $S_3$ summed up to a certain $l_3$, for FWHM beam-sizes 
of $7^\circ$, $13'$, and $5'\hspace{-2.5pt}.5$. 
These values correspond to beam-sizes of COBE, MAP, and Planck experiments,
respectively. Figure~\ref{fig:skewness} also 
plots the infinitesimal thin-beam case.
MAP, Planck, and the ideal experiments measure very similar $S_3$
one another, despite the fact that Planck and
the ideal experiments can use much more number of modes than MAP. 
The reason is as follows.
Looking at equation (\ref{eq:skewness}), one finds that $S_3$
is the linear integration of $b_{l_1l_2l_3}$ over $l_i$.
Thus, integrating oscillations in $b_{l_1l_2l_3}^{primary}$ around zero 
(see figure~\ref{fig:bispectrum}) damps the non-Gaussian signal in small 
angular scales, $l\gtrsim 300$.
Since the COBE scale is basically dominated by the Sachs--Wolfe effect, 
no oscillation, the cancellation affects $S_3$ less significantly 
than in MAP and Planck scales, while Planck suffers from severe 
cancellation in small angular scales.
Even Planck and the ideal experiments measure 
the same amount of $S_3$ as MAP does.
As a result, measured $S_3$ almost saturates at the MAP resolution
scale, $l\sim 500$.

We conclude this section by noting that 
when we can calculate the expected form of the bispectrum, then it is
a ``matched filter'' for detecting the non-Gaussianity in the data,
and thus much more powerful tool than the skewness in which the information
is lost through the coarse-graining.

\section{Secondary sources of the CMB bispectrum}
\label{sec:secondary}

Even if the CMB bispectrum were significantly detected in the CMB map, 
the origin would not necessarily be primordial, but rather
would be various foregrounds such as the Sunyaev--Zel'dovich effect\cite{ZS69}
(hereafter SZ), 
the weak lensing effect, extragalactic radio sources, 
and so on. In order to isolate the primordial origin from others, 
we have to know the accurate form of bispectra produced by 
the foregrounds.

\subsection{Coupling between the weak lensing and the 
Sunyaev--Zel'dovich effects}

The coupling between the SZ and the weak lensing
effects would produce an observable effect in the
bispectrum\cite{GS99,CH00}. 
The CMB temperature field including the SZ and the lensing effects 
is expanded as
\begin{equation}
  \label{eq:lenscouple}
  \frac{\Delta T(\hat{\mathbf n})}T
  = \frac{\Delta T^P\left(
                          \hat{\mathbf n}+\nabla\Theta(\hat{\mathbf n})
                    \right)}T
   +\frac{\Delta T^{SZ}(\hat{\mathbf n})}T
  \approx \frac{\Delta T^P(\hat{\mathbf n})}T
  +\nabla\left(\frac{\Delta T^P(\hat{\mathbf n})}T\right)\cdot
   \nabla\Theta(\hat{\mathbf n})
  +\frac{\Delta T^{SZ}(\hat{\mathbf n})}T,
\end{equation}
where $P$ denotes the primary anisotropy, 
$\Theta(\hat{\mathbf n})$ is the lensing potential:
\begin{equation}
 \Theta(\hat{\mathbf n})
  \equiv
  -2\int_0^{r_*} dr \frac{r_*-r}{rr_*}
  \Phi(r,\hat{\mathbf n}r),
\end{equation}
and $SZ$ denotes the SZ effect:
\begin{equation}
  \label{eq:sz}
  \frac{\Delta T^{SZ}(\hat{\mathbf n})}T
  =
  y(\hat{\mathbf n})j_\nu,
\end{equation}
where $j_\nu$ is the spectral function of the SZ effect\cite{ZS69}.
$y(\hat{\mathbf n})$ is the Compton $y$-parameter given by
\begin{equation}
  \label{eq:yparam}
  y(\hat{\mathbf n}) \equiv y_0\int \frac{dr}{r_*} 
  \frac{T_\rho(r,\hat{\mathbf n}r)}{\overline{T}_{\rho0}} a^{-2}(r),
\end{equation}
where
\begin{equation}
  \label{eq:y0}
  y_0\equiv \frac{\sigma_T \overline{\rho}_{gas0} k_B 
            \overline{T}_{\rho0} r_*}{\mu_e m_p m_e c^2} 
     = 4.3\times 10^{-4}\mu_e^{-1}\left(\Omega_b h^2\right) 
       \left(\frac{k_B \overline{T}_{\rho0}}{1~{\rm keV}}\right)
       \left(\frac{r_*}{10~{\rm Gpc}}\right).
\end{equation}
$T_\rho\equiv \rho_{gas} T_e/\overline{\rho}_{gas}$ is the 
electron temperature weighted by the gas mass density, the overline denotes
the volume average, and the subscript 0 means the present epoch.
We adopt $\mu_e^{-1}=0.88$, where 
$\mu_e^{-1}\equiv n_e/(\rho_{gas}/m_p)$ is the number of electrons 
per proton mass in fully ionized medium.
Other quantities have their usual meanings.

Transforming equation (\ref{eq:lenscouple}) into harmonic space, 
\begin{eqnarray}
  \nonumber
  a_{lm}&=&
  a_{lm}^P
  +\sum_{l'm'}\sum_{l''m''}(-1)^m
  {\cal G}_{l l' l''}^{-m m' m''}
  \frac{l'(l'+1)-l(l+1)+l''(l''+1)}2
  a_{l'm'}^P\Theta_{l''m''}
  +a_{lm}^{SZ}\\
  \label{eq:almlens}
  &=&
  a_{lm}^P
  +\sum_{l'm'}\sum_{l''m''}(-1)^{m+m'+m''}
  {\cal G}_{l l' l''}^{-m m' m''}
  \frac{l'(l'+1)-l(l+1)+l''(l''+1)}2
  a_{l'-m'}^{P*}\Theta^*_{l''-m''}
  +a_{lm}^{SZ},
\end{eqnarray}
where ${\cal G}_{l_1l_2l_3}^{m_1m_2m_3}$ is the Gaunt integral
(Eq.(\ref{eq:gaunt})).
Substituting equation (\ref{eq:almlens}) into (\ref{eq:blllmmm}),
and using the identity ${\cal G}_{l_1l_2l_3}^{-m_1-m_2-m_3}
={\cal G}_{l_1l_2l_3}^{m_1m_2m_3}$, we obtain the bispectrum,
\begin{equation}
  \label{eq:szlensbispec}
  B_{l_1l_2l_3}^{m_1m_2m_3}
  ={\cal G}_{l_1 l_2 l_3}^{m_1 m_2 m_3}
  \left[
  \frac{l_1(l_1+1)-l_2(l_2+1)+l_3(l_3+1)}2
  C_{l_1}^P \left<\Theta^*_{l_3m_3} a_{l_3m_3}^{SZ}\right>
  + \mbox{5 permutations}
  \right].
\end{equation}
The form of equation (\ref{eq:func}) is confirmed, and then
the reduced bispectrum $b_{l_1l_2l_3}^{sz-lens}$ 
includes terms in the square bracket. 

The cross-correlation power spectrum of
the lensing and the SZ effects, 
$\left<\Theta^*_{lm} a_{lm}^{SZ}\right>$, 
appearing in equation (\ref{eq:szlensbispec})
was first derived by Goldberg and Spergel\cite{GS99}. 
They assumed the linear pressure bias model proposed by 
Persi et al.\cite{PSCO95}:
$T_\rho=\overline{T}_\rho b_{gas} \delta$,
and the mean temperature evolution of
$\overline{T}_\rho\simeq \overline{T}_{\rho0}(1+z)^{-1}$ 
for $z<2$ as roughly suggested by recent hydrodynamic 
simulations\cite{CO99,RKSP00,SWH00}.
Then they derived
\begin{equation}
  \label{eq:blsz}
  \left<\Theta^*_{lm} a_{lm}^{SZ}\right>
  \simeq
  -j_\nu\frac{4y_0b_{gas}l^2}{3\Omega_mH_0^2}
  \int_0^{z_*} dz \frac{dr}{dz}D^2(z)(1+z)^2
  \frac{r_*-r(z)}{r_*^2r^5(z)}
  P_\Phi\left(k=\frac{l}{r(z)}\right),
\end{equation}
where $D(z)$ is the linear growth factor.
Simulations without non-gravitational heating\cite{RKSP00,SWH00} 
suggest that $\overline{T}_{\rho0}\sim 0.2-0.4~{\rm keV}$ and 
$b_{gas}\sim 5-10$, and similar numbers are 
obtained by analytic estimations\cite{RKSP00,ZP00}.
In this pressure bias model, free parameters except cosmological 
parameters are $\overline{T}_{\rho0}$ and $b_{gas}$. 
However, both actually depend on cosmological models\cite{RKSP00}.
Since 
$l^3\left<\Theta^*_{lm}a_{lm}^{SZ}\right>\sim 
2\times 10^{-10}j_\nu \overline{T}_{\rho0}b_{gas}$
\cite{GS99,CH00} and $l^2C_l^P\sim 6\times 10^{-10}$,
\begin{equation}
  \label{eq:orderestszlens}
  b_{lll}^{sz-lens}\sim 
  l^{-3}
  \left[
  \left(l^2C_l^P\right)
  \left(l^3\left<\Theta^*_{lm} a_{lm}^{SZ}\right>\right)\times 5/2\right]
  \sim 
  l^{-3}\times
  3\times 10^{-19}j_\nu\overline{T}_{\rho0}b_{gas},
\end{equation}
where $\overline{T}_{\rho0}$ is in units of 1~keV, and
$b_{l_1l_2l_3}=
B_{l_1l_2l_3}^{m_1m_2m_3}\left({\cal
G}_{l_1l_2l_3}^{m_1m_2m_3}\right)^{-1}$ (Eq.(\ref{eq:func}))
is the reduced bispectrum.
Thus, comparing this to equation (\ref{eq:orderest}), we obtain
\begin{equation}
  \label{eq:ordercomp}
  \frac{b_{lll}^{primary}}{b_{lll}^{sz-lens}}  
  \sim l^{-1}\times 10
  \left(\frac{f_{NL}}{j_\nu\overline{T}_{\rho0}b_{gas}}\right).
\end{equation}
This estimate suggests that the primary bispectrum is overwhelmed by
the SZ--lensing bispectrum in small angular scales.
This is why we have to separate the primary from the SZ--lensing effect.

\subsection{Extragalactic radio and infrared sources}

The bispectrum from extragalactic radio and infrared sources 
whose fluxes $F$ are less than a certain detection threshold $F_d$ is 
relatively simple to estimate, 
when they are assumed to be Poisson distributed.
Since the Poisson distribution has the white noise spectrum, 
the reduced bispectrum (Eq.(\ref{eq:func})) is constant,
$b_{l_1l_2l_3}^{ps}=b^{ps}={\rm constant}$, then we obtain
\begin{equation}
  \label{eq:pointsource}
  B_{l_1l_2l_3}^{m_1m_2m_3}
  ={\cal G}_{l_l1_2l_3}^{m_1m_2m_3} b^{ps},
\end{equation}
where 
\begin{equation}
  \label{eq:Bps}
  b^{ps}(<F_d)
  \equiv g^3(x)
  \int_0^{F_d} dF F^3 \frac{dn}{dF}
  =
  g^3(x)\frac{n(>F_d)}{3-\beta}F_d^3.
\end{equation}
The assumption of the Poisson distribution is fairly good approximation
as found by Toffolatti et al.\cite{tof98}.
$dn/dF$ is the differential source count
per unit solid angle, and 
$n(>F_d)\equiv \int_{F_d}^\infty dF (dn/dF)$. 
The power law count, $dn/dF\propto F^{-\beta-1}$ with $\beta<2$,
has been assumed.  $x\equiv h\nu/k_B T \simeq 
(\nu/56.80~{\rm GHz})(T/2.726~{\rm K})^{-1}$, and
\begin{equation}
  \label{eq:gx}
  g(x)\equiv
  2\frac{(hc)^2}{(k_B T)^3}\left(\frac{\sinh x/2}{x^2}\right)^2
  \simeq
  \frac1{67.55~{\rm MJy~sr^{-1}}}\left(\frac{T}{2.726~{\rm K}}\right)^{-3}
  \left(\frac{\sinh x/2}{x^2}\right)^2.
\end{equation}
$b^{ps}$ is otherwise written in terms of the Poisson angular 
power spectrum $C^{ps}$:
\begin{equation}
  \label{eq:Cps}
  C^{ps}(<F_d)
  \equiv g^2(x)
  \int_0^{F_d} dF F^2 \frac{dn}{dF}
  =
  g^2(x)\frac{n(>F_d)}{2-\beta}F_d^2,
\end{equation}
as
\begin{equation}
  \label{eq:Bps*}
  b^{ps}(<F_d)
  = \frac{(2-\beta)^{3/2}}{3-\beta}\left[n(>F_d)\right]^{-1/2}
  \left[C^{ps}(<F_d)\right]^{3/2}.
\end{equation}

Toffolatti et al.\cite{tof98} estimated 
$n(>F_d)\sim 300~{\rm sr^{-1}}$ for $F_d\sim 0.2~{\rm Jy}$ at
217~GHz. This $F_d$ corresponds to $5\sigma$ detection
threshold for Planck experiment at 217~GHz.
Refregier, Spergel and Herbig\cite{RSH00} extrapolated Toffolatti et
al.'s estimation to 94~GHz, and obtained 
$n(>F_d)\sim 7~{\rm sr^{-1}}$ for $F_d\sim 2~{\rm Jy}$,
which corresponds to MAP $5\sigma$ threshold.
These values yield
\begin{eqnarray}
  \label{eq:cl90ghz}
  C^{ps}(90~{\rm GHz},<2~{\rm Jy})&\sim& 2\times 10^{-16},\\
  \label{eq:cl217ghz}
  C^{ps}(217~{\rm GHz},<0.2~{\rm Jy})&\sim& 1\times 10^{-17}.
\end{eqnarray}
Thus, rough estimates for $b^{ps}$ are
\begin{eqnarray}
  \label{eq:bl90ghz}
  b^{ps}(90~{\rm GHz},<2~{\rm Jy})&\sim& 2\times 10^{-25},\\
  \label{eq:bl217ghz}
  b^{ps}(217~{\rm GHz},<0.2~{\rm Jy})&\sim& 5\times 10^{-28}.
\end{eqnarray}
While we assumed the Euclidean source count ($\beta=3/2$) 
here for definiteness, this does not affect an order of magnitude
estimates above.
Since the primary reduced bispectrum $\propto l^{-4}$ (Eq.(\ref{eq:orderest}))
and the SZ--lensing reduced bispectrum $\propto l^{-3}$
(Eq.(\ref{eq:orderestszlens})), the Poisson bispectrum rapidly
becomes to dominate the total bispectrum in small angular scales,
\begin{eqnarray}
  \label{eq:ordercomp*}
  \frac{b_{lll}^{primary}}{b^{ps}}
  &\sim& l^{-4}\times 10^7
  \left(\frac{f_{NL}}{b^{ps}/10^{-25}}\right),\\
  \label{eq:ordercomp**}
  \frac{b_{lll}^{sz-lens}}
       {b^{ps}}
  &\sim& l^{-3}\times 10^6
  \left(\frac{j_\nu\overline{T}_{\rho0}b_{gas}}
             {b^{ps}/10^{-25}}\right).
\end{eqnarray}
For example, the SZ--lensing bispectrum measured by 
MAP experiment is overwhelmed by point sources at $l\gtrsim 100$.

\section{Measuring Bispectra}
\label{sec:measure}

\subsection{Fisher matrix}


We shall discuss the detectability of 
CMB experiments to the primary non-Gaussianity in the bispectrum.
We also need to separate it from secondary bispectra.
Suppose that we try to fit the observed bispectrum $B_{l_1l_2l_3}^{obs}$
by theoretically calculated bispectra which include 
both primary and secondary sources.
Then we minimize $\chi^2$ defined by
\begin{equation}
  \label{eq:chisq}
  \chi^2
  \equiv 
  \sum_{2\leq l_1\leq l_2\leq l_3}
  \frac{\left(B_{l_1l_2l_3}^{obs}
	     -\sum_i A_i B^{(i)}_{l_1l_2l_3}\right)^2}
  {\sigma^2_{l_1l_2l_3}},
\end{equation}
where $i$ denotes a component such as
the primary, the SZ and lensing effects, extragalactic sources, and so on. 
Unobservable modes $l=0$ and $1$ are removed.
In case that the non-Gaussianity is small, the cosmic variance
of the bispectrum
is given by the six-point function of $a_{lm}$\cite{Luo94,heavens98}.
The variance of $B_{l_1l_2l_3}$ is then 
calculated as \cite{SG99,GM00}
\begin{equation}
  \sigma^2_{l_1l_2l_3}
  \equiv \left<B_{l_1l_2l_3}^2\right>-\left<B_{l_1l_2l_3}\right>^2
  \approx
  {\cal C}_{l_1}{\cal C}_{l_2}{\cal C}_{l_3}\Delta_{l_1l_2l_3},
\end{equation}
where $\Delta_{l_1l_2l_3}$ takes values 1, 2, and 6 
for cases of that all $l$'s are different,
 two of them are same, and all are same, respectively. 
${\cal C}_l\equiv C_l+C_l^N$ is the total CMB angular power spectrum,
which includes the power spectrum of the detector noise $C_l^N$.
$C_l^N$ is calculated analytically using the formula derived by 
Knox\cite{Knox95}
with the noise characteristics of the relevant 
experiments.
We do not include $C_l$ from secondary sources, as they are
totally subdominant compared with the primary $C_l$ and $C_l^N$
for relevant experiments.
For example, inclusion of $C_l$ from extragalactic sources
(Eqs.(\ref{eq:cl90ghz}) or (\ref{eq:cl217ghz})) changes our 
results less than 10\%.

Taking $\partial\chi^2/\partial A_i=0$, we obtain the normal equation,
\begin{equation}
  \label{eq:fiseq}
  \sum_j
  \left[\sum_{2\leq l_1\leq l_2\leq l_3}
        \frac{B_{l_1l_2l_3}^{(i)}B_{l_1l_2l_3}^{(j)}}{\sigma_{l_1l_2l_3}^2}
  \right]A_j
  =
  \sum_{2\leq l_1\leq l_2\leq l_3}
        \frac{B_{l_1l_2l_3}^{obs}B_{l_1l_2l_3}^{(i)}}{\sigma_{l_1l_2l_3}^2}.
\end{equation}
Thus, we define the Fisher matrix $F_{ij}$ as
\begin{equation}
  \label{eq:fis}
  F_{ij}\equiv 
  \sum_{2\leq l_1\leq l_2\leq l_3}
  \frac{B_{l_1l_2l_3}^{(i)}B_{l_1l_2l_3}^{(j)}}{\sigma_{l_1l_2l_3}^2}
  =
  \frac{2}{\pi}\sum_{2\leq l_1\leq l_2\leq l_3}
  \left(l_1+\frac12\right)\left(l_2+\frac12\right)\left(l_3+\frac12\right)
  \left(\begin{array}{ccc}l_1&l_2&l_3\\0&0&0\end{array}\right)^2
  \frac{b_{l_1l_2l_3}^{(i)}b_{l_1l_2l_3}^{(j)}}{\sigma_{l_1l_2l_3}^2},
\end{equation}
where we have used equation (\ref{eq:wigner*}) to replace $B_{l_1l_2l_3}$ by
the reduced bispectrum $b_{l_1l_2l_3}$ (see Eq.(\ref{eq:func}) for
definition). 
Since the covariance matrix of $A_i$ is $F_{ij}^{-1}$,
we define the signal-to-noise ratio $(S/N)_i$ for 
a component $i$, the correlation coefficient $r_{ij}$ between 
different components $i$ and $j$, and
the degradation parameter $d_i$ of $(S/N)_i$ due to $r_{ij}$ as
\begin{eqnarray}
  \label{eq:sn}
  \left(\frac{S}{N}\right)_i &\equiv& \frac1{\sqrt{F_{ii}^{-1}}},\\
  \label{eq:r}
  r_{ij}&\equiv& 
  \frac{F_{ij}^{-1}}{\sqrt{F^{-1}_{ii}F^{-1}_{jj}}},\\
  \label{eq:d}
  d_i&\equiv& F_{ii}F_{ii}^{-1}.
\end{eqnarray}
Note that $r_{ij}$ does not depend on amplitudes of bispectra,
but shapes. $d_i$ is defined so as $d_i=1$ for zero degradation,
while $d_i>1$ for degraded $(S/N)_i$.
Spergel and Goldberg\cite{SG99} and Cooray and Hu\cite{CH00}
considered the diagonal component of $F_{ij}^{-1}$,
while we study all components in order to discuss 
the separatability between various bispectra.

An order of magnitude estimation of $S/N$ as a function of 
a certain angular resolution $l$ is possible as follows.
Since the number of modes contributing to $S/N$ increases as 
$l^{3/2}$ and 
$l^3\left(\begin{array}{ccc}l&l&l\\0&0&0\end{array}\right)^2
\sim 0.36\times l$, we estimate $(S/N)_i\sim (F_{ii})^{1/2}$ as
\begin{equation}
  \label{eq:snorder}
  \left(\frac{S}{N}\right)_i
  \sim 
  \frac1{3\pi} l^{3/2}
  \times l^{3/2}
  \left|
  \left(\begin{array}{ccc}l&l&l\\0&0&0\end{array}\right)
  \right|\times
   \frac{l^3b_{lll}^{(i)}}
        {(l^2 C_l)^{3/2}}
  \sim l^5b_{lll}^{(i)}\times 4\times 10^{12},
\end{equation}
where we have used $l^2C_l\sim 6\times 10^{-10}$.

Table~\ref{tab:fis} and \ref{tab:invfis} tabulate all components 
of $F_{ij}$ and $F_{ij}^{-1}$,
respectively. Table~\ref{tab:sn} summarizes $(S/N)_i$, while 
table~\ref{tab:corr} tabulates
$d_i$ in the diagonal, and $r_{ij}$ in the off-diagonal parts.

\subsection{Measuring primary bispectrum}

\begin{figure}
\begin{center}
    \leavevmode\epsfxsize=9cm \epsfbox{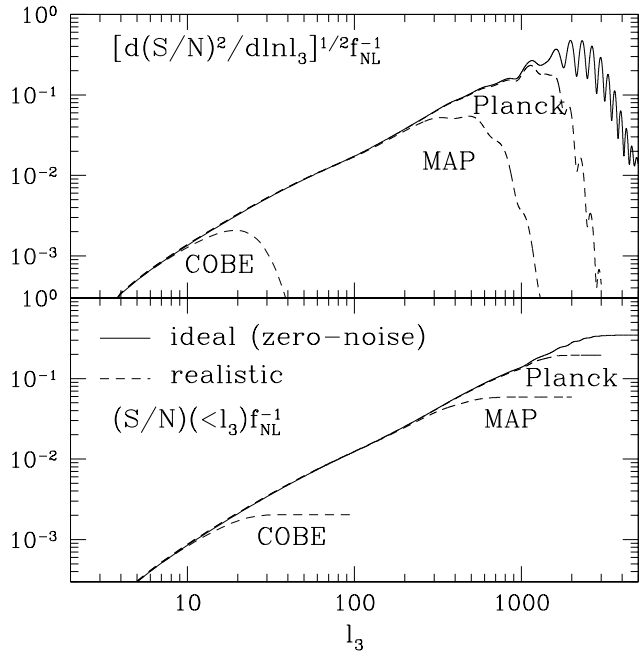}
\end{center}
\caption{
The predictions of the signal-to-noise ratio, $S/N$, 
for COBE, MAP, and Planck experiments (see Eq.(\ref{eq:sn})).
The differential $S/N$ at $\ln l_3$ interval is shown in the upper
panel, while the cumulative $S/N$ up to a certain $l_3$ is shown
in the lower panel. Both are in units of $f_{NL}$.
Solid line represents the zero-noise ideal experiment, 
while dotted lines show
the realistic experiments mentioned above.
The total $(S/N)f^{-1}_{NL}$ 
are $1.7\times 10^{-3}$, $5.8\times 10^{-2}$, and $0.19$
for COBE, MAP, and Planck experiments, respectively.
}
\label{fig:fis11}
\end{figure}

Figure~\ref{fig:fis11} shows the numerical results of
differential $S/N$ for the primary bispectrum at $\ln l_3$ interval,
$\left[d(S/N)^2/d\ln l_3\right]^{1/2}f_{NL}^{-1}$, in the upper panel, and 
$(S/N)(<l_3)f_{NL}^{-1}$, which is $S/N$ summed up to a certain $l_3$, 
in the lower panel. 
The detector noises $C_l^N$ have been computed for 
COBE 4-yr map\cite{cobe96}, for MAP 90 GHz channel, 
and for Planck 217 GHz channel, but
the effect of limited sky coverage is neglected.
Figure~\ref{fig:fis11} also shows results for the ideal experiment 
with no noise: $C_l^N=0$.
Both $\left[d(S/N)^2/d\ln l_3\right]^{1/2}$ and 
$(S/N)(<l_3)$ are monotonically 
increasing function with $l_3$ as
roughly $\propto l_3$ up to $l_3\sim 2000$ for the ideal experiment.

Beyond $l_3\sim 2000$, an enhancement of the damping tail in $C_l$ 
because of the weak lensing effect\cite{Seljak96} stops 
$\left[d(S/N)^2/d\ln l_3\right]^{1/2}$ and then $(S/N)(<l_3)$
increasing. 
This leads to an important constraint on the observation; 
even for the ideal noise-free and the 
infinitesimal thin-beam experiment, 
there is an upper limit on the value of $S/N\lesssim 0.3f_{NL}$.
For a given realistic experiment, 
$\left[d(S/N)^2/d\ln l_3\right]^{1/2}$ has a maximum at 
a scale near the beam-size.

The total $(S/N)f^{-1}_{NL}$ are
$1.7\times 10^{-3}$, $5.8\times 10^{-2}$, and $0.19$
for COBE, MAP and Planck experiments, respectively (see table~\ref{tab:sn}).
In order to obtain $S/N>1$, 
therefore, we need $f_{NL}>600,20$, and $5$ for each corresponding 
experiment, while the ideal experiment requires $f_{NL}>3$
(see table~\ref{tab:fnl}).
These values are also roughly obtained by 
substituting equation (\ref{eq:orderest}) into (\ref{eq:snorder}),
\begin{equation}
  \label{eq:snorderprim}
  \left(\frac{S}{N}\right)_{primary}
  \sim 
  l \times 10^{-4}f_{NL}.
\end{equation}
 
The degradation parameters $d_{primary}$ 
are 1.46, 1.01, and 1.00 for COBE, 
MAP, and Planck experiments, respectively (see table~\ref{tab:corr}).
This means that MAP and Planck experiments will separate the 
primary bispectrum from others at 1\% or better accuracies.
Since the primary and other secondary sources change 
monotonically in the COBE angular scales, COBE cannot 
discriminate between them very well. 
In the MAP and Planck scales, however, the primary bispectrum
starts oscillating around zero, 
and then is well separated in shape 
from other secondaries, as the secondaries do not oscillate.
This is good news for the forthcoming high angular
resolution CMB experiments.

\subsection{Measuring secondary bispectra}

The signal-to-noises for measuring the SZ--lensing bispectrum 
$(S/N)_{sz-lens}$ in units of $\left|j_\nu\right| \overline{T}_{\rho0}b_{gas}$
are $1.8\times 10^{-4}$, $0.34$, and $6.2$ for COBE, MAP, and Planck
experiments, respectively (see table~\ref{tab:sn}).
$\overline{T}_{\rho0}$ is in units of 1~keV.
Using equations (\ref{eq:snorder}) and (\ref{eq:orderestszlens}), 
we roughly estimate $(S/N)_{sz-lens}$ as
\begin{equation}
  \label{eq:snorderszlens}
  \left(\frac{S}{N}\right)_{sz-lens}
  \sim 
  l^2 \times 10^{-6}\left|j_\nu\right| \overline{T}_{\rho0}b_{gas}.
\end{equation}
Thus, $(S/N)_{sz-lens}$ increases with the angular resolution
more rapidly than the primary bispectrum (see
Eq.(\ref{eq:snorderprim})).
Since $\left|j_\nu\right| \overline{T}_{\rho0}b_{gas}$ should be
an order of unity, COBE and MAP would not be expected to detect the 
SZ--lensing bispectrum; however, Planck would be sensitive enough 
to detect, depending on the frequency, i.e., a value of $j_\nu$.
For example, 217~GHz is totally insensitive to the SZ effect as
$j_\nu\sim 0$, while $j_\nu=-2$ in the Rayleigh--Jeans regime.

The degradation parameters $d_{sz-lens}$ are
3.89, 1.16, and 1.00 for COBE, MAP, and Planck experiments,
respectively (see table~\ref{tab:corr}).
Thus, Planck will separate the SZ--lensing bispectrum from other 
effects.
Note that $(S/N)_{sz-lens}$ values 
must be an order of magnitude estimation,
as our cosmological model is the 
COBE normalized SCDM yielding $\sigma_8=1.2$.
Since this $\sigma_8$ is about a factor of 2 greater than the cluster
normalization with $\Omega_m=1$, and $20\%$ greater than
the normalization with $\Omega_m=0.3$\cite{KS97}.
Thus, this factor tends to overestimate 
$\left<\Theta^*_{lm} a_{lm}^{SZ}\right>$ (Eq.(\ref{eq:blsz})) by a 
factor of several. 
On the other hand, using the linear power spectrum for $P_\Phi(k)$ 
rather than the non-linear power spectrum tends to underestimate the effect
by a factor of several at $l\sim 3000$\cite{CH00}.
However, our main goal is to discriminate between 
shapes of various bispectra, not amplitudes, so that this factor 
does not affect our conclusion on the degradation parameters $d_i$.

For the extragalactic radio and infrared sources, we estimated 
the signal-to-noises as $5.7\times 10^{-7}(b^{ps}/10^{-25})$, 
$2.2(b^{ps}/10^{-25})$, and $52(b^{ps}/10^{-27})$ for 
COBE, MAP, and Planck experiments, respectively (see
table~\ref{tab:sn}),
and the degradation parameters $d_{ps}$ are
3.45, 1.14, and 1.00 (see table~\ref{tab:corr}).
Since
\begin{equation}
  \left(\frac{S}{N}\right)_{ps}
  \sim l^5\times 10^{-13}\left(\frac{b^{ps}}{10^{-25}}\right),
\end{equation}
from equation (\ref{eq:snorder}),
$S/N$ of the bispectrum from point sources increases 
very rapidly with the angular resolution.
Our estimate that MAP will detect the bispectrum from point sources
is consistent with the results found by 
Refregier, Spergel and Herbig\cite{RSH00}. Although MAP cannot separate 
the Poisson bispectrum from the SZ--lensing bispectrum very well (see 
$r_{ij}$ in table~\ref{tab:corr}), it would not matter as 
the SZ--lensing bispectrum would be too small to be measured by MAP.
Planck will do an excellent job on separating all kinds of bispectra,
at least including the primary signal, SZ--lensing coupling, 
and extragalactic point sources, on the basis of the shape difference.

\subsection{Measuring primary skewness}

For the skewness, we define $S/N$ as
\begin{equation}
  \label{eq:skew_sn}
  \left(\frac{S}{N}\right)^2\equiv
  \frac{S_3^2}{\sigma^2_{S_3}},
\end{equation}
where the variance is\cite{Srednicki93}
\begin{eqnarray}
  \nonumber
  \sigma_{S_3}^2
  &\equiv& \left<\left(S_3\right)^2\right> =
  6\int_{-1}^{1}\frac{d\cos\theta}2 \left[{\cal C}(\theta)\right]^3\\
  \nonumber
  &=&
  6\sum_{l_1l_2l_3}
  \frac{\left(2l_1+1\right)\left(2l_2+1\right)\left(2l_3+1\right)}
        {(4\pi)^3}
  \left(
  \begin{array}{ccc}
  l_1 & l_2 & l_3 \\ 0 & 0 & 0 
  \end{array}
  \right)^2
   {\cal C}_{l_1}{\cal C}_{l_2}{\cal C}_{l_3}
   W^2_{l_1}W^2_{l_2}W^2_{l_3}\\
  \label{eq:skewvar}
  &=&
  \frac{9}{2\pi^3}\sum_{2\leq l_1\leq l_2\leq l_3}
  \left(l_1+\frac12\right)\left(l_2+\frac12\right)\left(l_3+\frac12\right)
  \left(
  \begin{array}{ccc}
  l_1 & l_2 & l_3 \\ 0 & 0 & 0 
  \end{array}
  \right)^2
   {\cal C}_{l_1}{\cal C}_{l_2}{\cal C}_{l_3}
   W^2_{l_1}W^2_{l_2}W^2_{l_3}.
\end{eqnarray}
In the last equality, we have used the symmetry of summed 
quantity with respect to indices (Eq.(\ref{eq:sumchange})), 
and removed unobservable modes $l=0$ and $1$.
Typically $\sigma_{S_3}\sim 10^{-15}$, as $\sigma_{S_3}\sim
\left[{\cal C}(0)\right]^{3/2}\sim 10^{-15}$, 
where ${\cal C}(\theta)$ is the temperature auto correlation function
including noise.
The lower panel of figure~\ref{fig:skewness} shows $\sigma_{S_3}(<l_3)$,  
which is $\sigma_{S_3}(<l_3)$ summed up to a certain $l_3$, for 
COBE, MAP, and Planck experiments as well as the ideal experiment.
Since ${\cal C}_{l}W^2_l= C_l e^{-l(l+1)\sigma^2_b} + w^{-1}$,
where $w^{-1}$ determines the white noise power spectrum of the detector
noise according to the Knox's formula\cite{Knox95},
the dominance of second term beyond the experimental angular resolution 
scale, $l\sim \sigma_b^{-1}$, keeps
$\sigma_{S_3}(<l_3)$ slightly increasing with $l_3$, while $S_3(<l_3)$ 
becomes constant beyond that (see the upper panel of 
figure~\ref{fig:skewness}). 
As a result, $S/N$ starts somewhat decreasing beyond the resolution.
We use the maximum $S/N$ for estimating the minimum
value of $f_{NL}$ needed to detect the primary $S_3$.
We find that 
$f_{NL}> 800$, 80, 70, and 60 for COBE, MAP, Planck, and the ideal
experiments, respectively, with all-sky coverage.

These $f_{NL}$ values are systematically larger than those needed to
detect $B_{l_1l_2l_3}$ by a factor of 1.3, 4, 14, and 20, respectively
(see table~\ref{tab:fnl}).
Higher the angular resolution, less sensitive measuring the primary $S_3$ 
than $B_{l_1l_2l_3}$.
This is because the cancellation
effect in smaller angular scales due to the oscillation of
$B_{l_1l_2l_3}$ damps $S_3$.

\begin{figure}
\begin{center}
    \leavevmode\epsfxsize=9cm \epsfbox{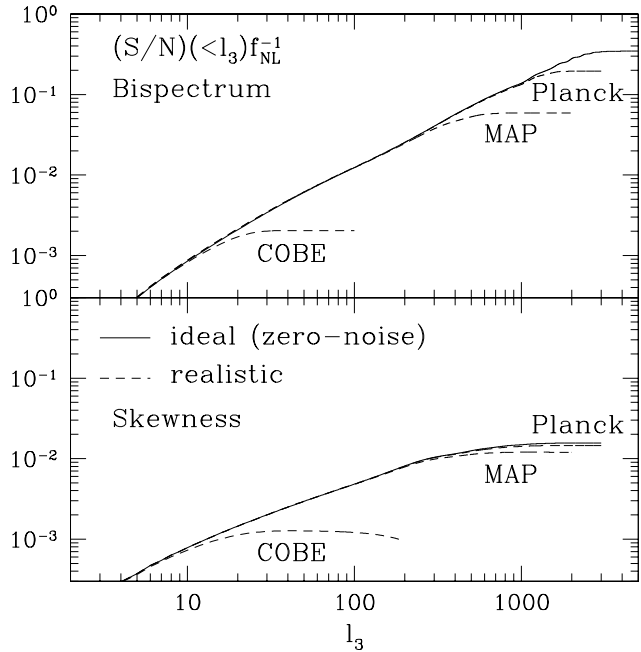}
\end{center}
\caption{
The comparison of the signal-to-noise ratio summed up to a certain
 $l_3$, $S/N(<l_3)$, for the detection of the bispectrum 
(upper panel; Eq.(\ref{eq:sn})) 
and the skewness (lower panel; Eq.(\ref{eq:skew_sn})) 
in units of $f_{NL}$
for COBE, MAP, and Planck experiments (dotted lines), and the 
ideal experiment (solid line).
See table~\ref{tab:fnl} for values of $f_{NL}$ in order to obtain $S/N>1$. 
}
\label{fig:sn}
\end{figure}

\section{Discussion and Conclusion}
\label{sec:discussion}


Using the full radiation transfer function , 
we have computed numerically the primary cosmic microwave background 
bispectrum (Eq.(\ref{eq:almspec})) and skewness 
(Eq.(\ref{eq:skewness})) down to arcminutes angular scales.
The primary bispectrum oscillates around zero
(figure~\ref{fig:bispectrum}), thus the primary 
skewness saturates at the MAP angular resolution scale,
$l\sim 500$ (figure~\ref{fig:skewness}).
We have introduced the ``reduced'' bispectrum defined by equation 
(\ref{eq:func}), and confirmed that this quantity is more useful to
describe the physical property of the bispectrum than the full
bispectrum (Eq.(\ref{eq:blllmmm})).

Figure~\ref{fig:sn} compares the expected signal-to-noise ratio 
for detecting the primary non-Gaussianity based on the bispectrum
(Eq.(\ref{eq:sn})) to that based on the skewness
(Eq.(\ref{eq:skew_sn})). 
It shows that the bispectrum is almost an order of magnitude more 
sensitive to the non-Gaussianity than the skewness.
We conclude that when we can compute the predicted form of the
bispectrum, it becomes a ``matched filter'' for detecting the 
non-Gaussianity in the data, and thus much more powerful tool than the 
skewness.
Table~\ref{tab:fnl} summarizes $f_{NL}$ required for detecting the primary 
non-Gaussianity using the bispectrum or the skewness 
with COBE, MAP, Planck, and the ideal experiments. 
This shows that even the ideal experiment 
needs $f_{NL}>3$ in order to detect the primary bispectrum.

We estimated the secondary bispectra from the coupling between the 
Sunyaev--Zel'dovich (SZ) and the weak lensing effects, and from 
the extragalactic radio and infrared sources.
Only Planck will detect the SZ--lensing bispectrum, while both MAP and 
Planck will detect the bispectrum from extragalactic point sources
(table~\ref{tab:sn}).

We also studied how well we can discriminate 
among the primary,
the SZ--lensing coupling, and the extragalactic point sources bispectra.
We found that MAP and Planck will separate the primary from other
secondary sources at 1\% or better accuracies. This conclusion is 
due to the presence of acoustic oscillation in the primary
bispectrum that does not appear in the secondary bispectra. 
The SZ--lensing coupling and the extragalactic sources are 
well separately measured by Planck experiment, although COBE and 
MAP cannot discriminate between them (table~\ref{tab:corr}).

Our arguments about the ability to discriminate among various bispectra
were fully based upon the shape difference, and thus did not take into 
account the spectral difference in the frequency space.
As pointed out by \cite{TE96,CHT00}, the multi-band
observation is so efficient to discriminate among the primary
signal and the other foreground contaminants for measuring the 
CMB anisotropy power spectrum.
Their scheme should be effective on the bispectrum as well, and
the accuracy of the foreground removal will be improved further. 
Thus, we expect that MAP and Planck will measure the primary
bispectrum separately from the foregrounds.

The simplest inflationary scenario
usually predicts small $f_{NL}$ $(\sim 10^{-2})$\cite{SB9091,GLMM94}, and
the second order perturbation theory yields $f_{NL}\sim 1$\cite{PC96}.
Thus, the significant detection of the primary bispectrum or the
skewness with any experiments means that the simplest inflationary 
scenario needs to be modified.
According to our results, if the reported
detections\cite{FMG98,Magueijo00} 
of the bispectrum in the COBE map were the cosmological origin, then
MAP and Planck would detect the primary bispectrum much more 
significantly.
Although Banday, Zaroubi and G\'orski\cite{BZG00} pointed out 
the one of those detections\cite{FMG98} could be accounted for by 
the experimental systematic effects of COBE,
the other\cite{Magueijo00} is claimed to be significant even after
removing such the systematics.

Although we have not discussed so far, 
the spatial distribution of emissions from interstellar dust
is a potential source of the microwave non-Gaussianity. 
Since it is very hard to estimate the bispectrum
analytically, the dust map compiled by Schlegel, 
Finkbeiner and Davis\cite{SFD98} could be used to estimate
the dust bispectrum. For example, we found that the dimensionless 
skewness parameter defined by 
$\left<(\Delta T)^3\right>/\left<(\Delta T)^2\right>^{3/2}$
is as large as 51. We used the publicly available 
HEALPix-formatted\cite{GHW98} $100~\mu{\rm m}$ map 
which contains 12,582,912 pixels without sky cut. The mean intensity
in the map was $14.8~{\rm MJy~sr^{-1}}$.
Of course, this skewness is 
largely an overestimate for the CMB measurement in reality; 
we need to cut a fraction of sky which contains the Galactic plane,
and then this will greatly reduce the non-Gaussianity.
Nevertheless, residual non-Gaussianity is still a source of the 
microwave bispectrum, and has to be taken into account.
Moreover, the form of the bispectrum measured in the dust map would reflect
the physics of interstellar dust, which is highly uncertain at present,
and thus studying the interstellar dust bispectrum would be 
challenging field.

\section*{Acknowledgments}

We would like to thank Naoshi Sugiyama and Licia Verde for useful comments,
and Uro$\check{\rm s}$ Seljak and Matias Zaldarriaga
for making their CMBFAST code publicly available.
E. K. acknowledges a fellowship
from the Japan Society for the Promotion of Science.
D. N. S. is partially supported by the MAP/MIDEX program.



\newpage
\begin{table}[h]
\caption{
The Fisher matrix $F_{ij}$ (Eq.(\ref{eq:fis})).
$i$ and $j$ denote components listed in the first row and the 
first column, respectively.
$\overline{T}_{\rho0}$ is in units of 1 keV, 
$b^{ps}_{25}\equiv b^{ps}/10^{-25}$, and
$b^{ps}_{27}\equiv b^{ps}/10^{-27}$.
}
\label{tab:fis}
\begin{center} 
  \begin{tabular}{cccc}
  COBE & primary & SZ--lensing & point sources \\
  \hline 
  primary &
  $4.2\times 10^{-6}~f_{NL}^2$ & 
  $-4.0\times 10^{-7}~f_{NL} j_\nu \overline{T}_{\rho0} b_{gas}$ &
  $-1.0\times 10^{-9}~f_{NL} b^{ps}_{25}$ \\
  SZ--lensing &
  & 
  $1.3\times 10^{-7}~(j_\nu \overline{T}_{\rho0} b_{gas})^2$ &
  $3.1\times 10^{-10}~j_\nu \overline{T}_{\rho0} b_{gas} b^{ps}_{25}$ \\
  point sources&
  & &
  $1.1\times 10^{-12}~(b^{ps}_{25})^2$ \\
  \hline
  MAP & & & \\
  \hline 
  primary &
  $3.4\times 10^{-3}~f_{NL}^2$ & 
  $2.6\times 10^{-3}~f_{NL} j_\nu \overline{T}_{\rho0} b_{gas}$ &
  $2.4\times 10^{-3}~f_{NL} b^{ps}_{25}$ \\
  SZ--lensing &
  & 
  $0.14~(j_\nu \overline{T}_{\rho0} b_{gas})^2$ &
  $0.31~j_\nu \overline{T}_{\rho0} b_{gas} b^{ps}_{25}$ \\
  point sources&
  & &
  $5.6~(b^{ps}_{25})^2$ \\
  \hline
  Planck & & & \\
  \hline 
  primary &
  $3.8\times 10^{-2}~f_{NL}^2$ & 
  $7.2\times 10^{-2}~f_{NL} j_\nu \overline{T}_{\rho0} b_{gas}$ &
  $1.6\times 10^{-2}~f_{NL} b^{ps}_{27}$ \\
  SZ--lensing &
  & 
  $39~(j_\nu \overline{T}_{\rho0} b_{gas})^2$ &
  $5.7~j_\nu \overline{T}_{\rho0} b_{gas} b^{ps}_{27}$ \\
  point sources&
  & &
  $2.7\times 10^3~(b^{ps}_{27})^2$
  \end{tabular}
\end{center}
\end{table}

\begin{table}[h]
\caption{
The inverse Fisher matrix $F_{ij}^{-1}$.
$i$ and $j$ denote components listed in the first row and the 
first column, respectively.
$\overline{T}_{\rho0}$ is in units of 1 keV, 
$b^{ps}_{25}\equiv b^{ps}/10^{-25}$, and
$b^{ps}_{27}\equiv b^{ps}/10^{-27}$.
}
\label{tab:invfis}
\begin{center} 
  \begin{tabular}{cccc}
  COBE & primary & SZ--lensing & point sources \\
  \hline 
  primary &
  $3.5\times 10^{5}~f_{NL}^{-2}$ & 
  $1.1\times 10^{6}~(f_{NL} j_\nu \overline{T}_{\rho0} b_{gas})^{-1}$ &
  $1.3\times 10^7~(f_{NL} b^{ps}_{25})^{-1}$ \\
  SZ--lensing &
  & 
  $3.1\times 10^7~(j_\nu \overline{T}_{\rho0} b_{gas})^{-2}$ &
  $-7.8\times 10^9~(j_\nu \overline{T}_{\rho0} b_{gas} b^{ps}_{25})^{-1}$ \\
  point sources&
  & &
  $3.1\times 10^{12}~(b^{ps}_{25})^{-2}$ \\
  \hline
  MAP & & & \\
  \hline 
  primary &
  $3.0\times 10^2~f_{NL}^{-2}$ & 
  $-6.1~(f_{NL} j_\nu \overline{T}_{\rho0} b_{gas})^{-1}$ &
  $0.21~(f_{NL} b^{ps}_{25})^{-1}$ \\
  SZ--lensing &
  & 
  $8.4~(j_\nu \overline{T}_{\rho0} b_{gas})^{-2}$ &
  $-0.46~(j_\nu \overline{T}_{\rho0} b_{gas} b^{ps}_{25})^{-1}$ \\
  point sources&
  & &
  $0.21~(b^{ps}_{25})^{-2}$ \\
  \hline
  Planck & & & \\
  \hline 
  primary &
  $26~f_{NL}^{-2}$ & 
  $-4.9\times 10^{-2}~(f_{NL} j_\nu \overline{T}_{\rho0} b_{gas})^{-1}$ &
  $-5.7\times 10^{-5}~(f_{NL} b^{ps}_{27})^{-1}$ \\
  SZ--lensing &
  & 
  $2.6\times 10^{-2}~(j_\nu \overline{T}_{\rho0} b_{gas})^{-2}$ &
  $-5.4\times 10^{-5}~(j_\nu \overline{T}_{\rho0} b_{gas} b^{ps}_{27})^{-1}$ \\
  point sources&
  & &
  $3.7\times 10^{-4}~(b^{ps}_{27})^{-2}$
  \end{tabular}
\end{center}
\end{table}

\begin{table}[h]
\caption{
The signal-to-noise ratio $(S/N)_i$ (Eq.(\ref{eq:sn})).
$i$ denotes a component listed in the first row.
$\overline{T}_{\rho0}$ is in units of 1 keV, 
$b^{ps}_{25}\equiv b^{ps}/10^{-25}$, and
$b^{ps}_{27}\equiv b^{ps}/10^{-27}$.
}
\label{tab:sn}
\begin{center} 
  \begin{tabular}{cccc}
       & primary & SZ--lensing & point sources \\
  \hline 
  COBE & $1.7\times 10^{-3}~f_{NL}$ 
       & $1.8\times 10^{-4}~\left|j_\nu\right| \overline{T}_{\rho0}b_{gas}$
       & $5.7\times 10^{-7}~b_{25}^{ps}$ \\
  MAP  & $5.8\times 10^{-2}~f_{NL}$ 
       & $0.34~\left|j_\nu\right| \overline{T}_{\rho0}b_{gas}$
       & $2.2~b_{25}^{ps}$ \\
  Planck & $0.19~f_{NL}$ 
         & $6.2~\left|j_\nu\right| \overline{T}_{\rho0}b_{gas}$
         & $52~b_{27}^{ps}$
  \end{tabular}
\end{center}
\end{table}

\begin{table}[h]
\caption{
The degradation parameter $d_i$ (Eq.(\ref{eq:d})) and correlation $r_{ij}$ 
(Eq.(\ref{eq:r})) matrix. 
$i$ and $j$ denote components listed in the first row and the 
first column, respectively.
$d_i$ for $i=j$, while $r_{ij}$ for $i\neq j$.
}
\label{tab:corr}
\begin{center} 
  \begin{tabular}{cccc}
  COBE & primary & SZ--lensing & point sources \\
  \hline 
  primary &
  $1.46$ & 
  $0.33~{\rm sgn}(j_\nu)$ &
  $1.6\times 10^{-2}$ \\
  SZ--lensing &
  & 
  $3.89$ &
  $-0.79~{\rm sgn}(j_\nu)$ \\
  point sources&
  & &
  $3.45$ \\
  \hline
  MAP & & & \\
  \hline 
  primary &
  $1.01$ & 
  $-0.12~{\rm sgn}(j_\nu)$ &
  $2.7\times 10^{-2}$ \\
  SZ--lensing &
  & 
  $1.16$ &
  $-0.35~{\rm sgn}(j_\nu)$ \\
  point sources&
  & &
  $1.14$ \\
  \hline
  Planck & & & \\
  \hline 
  primary &
  $1.00$ & 
  $-5.9\times 10^{-2}~{\rm sgn}(j_\nu)$ &
  $-5.8\times 10^{-4}$ \\
  SZ--lensing &
  & 
  $1.00$ &
  $-1.8\times 10^{-2}~{\rm sgn}(j_\nu)$ \\
  point sources&
  & &
  $1.00$
  \end{tabular}
\end{center}
\end{table}

\begin{table}[h]
\caption{
The minimum non-linear coupling constant $f_{NL}$ required to detect
the primary non-Gaussianity by the bispectrum and the skewness
with the signal-to-noise ratio of $>1$. These estimates include
the effects of cosmic variance, detector noise, and foreground sources.
}
\label{tab:fnl}
\begin{center} 
  \begin{tabular}{ccc}
  Experiments & $f_{NL}$ (Bispectrum) & $f_{NL}$ (Skewness) \\
  \hline 
  COBE   & 600 & 800 \\
  MAP    & 20  & 80 \\
  Planck & 5   & 70 \\
  Ideal  & 3   & 60 \\
  \end{tabular}
\end{center}
\end{table}

\end{document}